\documentclass{svjour3}
\usepackage{amsmath}
\usepackage{amssymb}
\usepackage{algorithm}
\usepackage{algpseudocode}

\newenvironment{Proof}{\begin{proof}}{\qed\end{proof}}

\renewcommand{\emptyset}{\varnothing}

\newcommand{\und}{\mathrm{und}}
\newcommand{\multiset}[2]{\ensuremath{\left(\kern-.3em\left(\genfrac{}{}{0pt}{}{#1}{#2}\right)\kern-.3em\right)}}

\titlerunning{Minimal Output Unstable Configurations in CRNs and CRDs}

\title{Minimal Output Unstable Configurations in Chemical Reaction Networks and Deciders}
\author{Robert Brijder}
\institute{Hasselt University and Transnational University of Limburg, Belgium \email{robert.brijder@uhasselt.be}}

\begin{document}

\maketitle

\begin{abstract}
We study the set of output stable configurations of chemical reaction deciders (CRDs). It turns out that CRDs with only bimolecular reactions (which are almost equivalent to population protocols) have a special structure that allows for an algorithm to efficiently compute their finite set of minimal output unstable configurations. As a consequence, a relatively large set of configurations may be efficiently checked for output stability.

We also provide a number of observations regarding the semilinearity result of Angluin et al.\ [Distrib.\ Comput., 2007] from the context of population protocols (which is a central result for output stable CRDs). In particular, we observe that the computation-friendly class of totally stable CRDs has equal expressive power as the larger class of output stable CRDs.
\end{abstract}

\section{Introduction}
In scenarios where the number of molecules in a chemical reaction network (CRN) is small, traditional continuous models for CRNs based on mass action kinetics are not suitable and one may need to consider discrete CRNs. In discrete CRNs, the number of molecules of each species is represented by a nonnegative integer and probabilities are assigned to each reaction. The computational power of discrete CRNs has been formally studied in \cite{DBLP:journals/nc/SoloveichikCWB08} (see also \cite{ProgrCRNs/winfree_solo}), where it is shown that Turing-universal computation is possible with arbitrary small (but nonzero) error probability. The implementability of arbitrary CRNs has been studied using strand displacement reactions as a primitive \cite{SimCRN/Soloveichik}. As observed in \cite{DBLP:journals/nc/SoloveichikCWB08}, discrete CRNs are similar to population protocols \cite{DBLP:journals/dc/AngluinADFP06,DBLP:journals/eatcs/AspnesR07} and results carry over from one domain to the other. Recent work related to CRNs include the calculus of chemical systems \cite{Plotkin/CalculusChemical}, the study of timing issues in CRNs \cite{DBLP:conf/soda/Doty14}, and the study of rate-independent continuous CRNs \cite{DBLP:conf/innovations/ChenDS14}. From now on we consider only discrete CRNs, and so we omit the adjective ``discrete''.

We continue in this paper the study of CRNs that has for each given input a deterministic output \cite{DBLP:conf/dna/ChenDS12}. Thus, we are concerned here with error-free computation and so probabilities are irrelevant and only reachability is important. A given input is accepted by such a ``deterministic'' CRN, or more precisely \emph{output stable chemical reaction decider} (CRD) \cite{DBLP:conf/dna/ChenDS12}, if at the end of the ``useful'' computation we obtain an accept configuration $c$, which is a configuration where at least one yes voter is present and none of the no voters (each species is marked by the CRD as either a yes or a no voter). Otherwise, the input is rejected and $c$ is a reject configuration, which is a configuration where at least one no voter is present and none of the yes voters. The configuration $c$ may still change, but it stays an accept configuration when $c$ is an accept configuration (and similar for reject). In this case $c$ is called \emph{output stable}.

In Section~\ref{sec:semilinear}, we provide a number of observations regarding the semilinearity result for population protocols of \cite{DBLP:journals/dc/AngluinADFP06,DBLP:conf/podc/AngluinAE06}. First we mention that this result has a small gap in its proof which is easily fixable, except for the corner case where the semilinear set contains the zero vector. Next, we define a stricter variant of the notion of output stable, called \emph{totally stable}. In contrast to output stable CRDs, totally stable CRDs eventually (completely) halt for every input. For totally stable CRDs it is computationally easy to determine when the computation has ended. We mention that the semilinearity result of \cite{DBLP:journals/dc/AngluinADFP06,DBLP:conf/podc/AngluinAE06} works also for totally stable CRDs, and consequently the class of totally stable CRDs has equal expressive power as the larger class of output stable CRDs.

CRNs are similar to Petri nets \cite{PetriNet/review/Pet1977} and vector addition systems (VASs) \cite{VASsKarpMiller}, see \cite{DBLP:journals/nc/SoloveichikCWB08}. However, Petri nets and VASs operate as ``generators'' where the computation starts in the given fixed starting configuration (called the initial marking) and one is (generally) interested in the reachable configurations. In contrast, a CRD is a decider where one is (generally) interested in determining the set of inputs that is accepted by the CRD. Despite these differences, various results concerning Petri nets and VASs can be carried over to CRDs.

In Section~\ref{sec:det_output_stable}, we take a closer look at the notion of output stable. First, using some well-known results for VASs, we show that determining whether or not a configuration is output stable for an output stable CRD is decidable. Next, we turn to bimolecular CRNs, i.e., CRNs where each reaction has two reactants and two products. It turns out that bimolecular CRDs provide a special structure on the set of output stable configurations. More precisely, it turns out that the set of minimal elements $M$ of the upward-closed set of output-unstable configurations may be efficiently determined for bimolecular CRDs, cf.\ Theorem~\ref{thm:complexity_alg} --- this is the main result of the paper. By efficiently determine we mean here that the fraction of time complexity divided by size of the output is relatively small (note that the size of the output is a naive lower bound on the time complexity). Given $M$, it is then computationally easy to determine if a given configuration $c$ is output stable. Consequently, the algorithm to determine $M$ provides an efficient method to test a relatively large number of configurations for output stability (the preprocessing cost to generate $M$ becomes smaller, relatively, when testing more configurations for output stability).

A preliminary conference version of this paper was presented at DNA~20 \cite{DBLP:conf/dna/Brijder14}.

\section{Chemical Reaction Networks and Deciders and Population Protocols} \label{sec:CRNS_recall}

\subsection{Chemical Reaction Networks} \label{ssec:crns}

The notation and terminology of this subsection and the next are similar as in \cite{DBLP:conf/dna/DotyH13}.

Let $\mathbb{N} = \{0,1,\ldots\}$. Let $\Lambda$ be a finite set. The set of vectors over $\mathbb{N}$ indexed by $\Lambda$ (i.e., the set of functions $\varphi: \Lambda \rightarrow \mathbb{N}$) is denoted by $\mathbb{N}^\Lambda$. For $x \in \mathbb{N}^\Lambda$, we define the \emph{size} of $x$, denoted by $\| x \|$, as $\sum_{i \in \Lambda} x(i)$. We denote the restriction of $x$ to $\Sigma \subseteq \Lambda$ by $x|_{\Sigma}$. For $x,y \in \mathbb{N}^\Lambda$ we write $x \leq y$ if and only if $x(i) \leq y(i)$ for all $i \in \Lambda$. For notational convenience we now also denote vectors in $\mathbb{N}^{\Lambda}$, which can be regarded as multisets, by their string representations. Thus we denote $c \in \mathbb{N}^{\Lambda}$ by the string $A_1^{c(A_1)}\cdots A_n^{c(A_n)}$ (or any permutation of these letters) where $\Lambda = \{A_1, \ldots, A_n\}$.

Let $\Lambda$ be a finite set. A \emph{reaction} $\alpha$ over $\Lambda$ is a tuple $(r,p)$ with $r,p \in \mathbb{N}^\Lambda$; $r$ and $p$ are called the \emph{reactants} and \emph{products} of $\alpha$, respectively. A reaction is commonly written in an additive fashion, where for example
$A+2B \to C$ denotes the reaction $(r,p)$ where $AB^2$ and $C$ are string representations for $r$ and $p$, respectively.
We say that $\alpha$ is \emph{mute} if $r=p$. We say that $\alpha$ is \emph{nonincreasing} if $\|r\| \geq \|p\|$ and \emph{bimolecular} if $\|r\| = \|p\| = 2$. A \emph{chemical reaction network} (CRN, for short) is a tuple $\mathcal{R} = (\Lambda, R)$ with $\Lambda$ a finite set and $R$ a finite set of reactions over $\Lambda$. The elements of $\Lambda$ are called the \emph{species} of $\mathcal{R}$. The elements of $\mathbb{N}^\Lambda$ are called the \emph{configurations} of $\mathcal{R}$. For a configuration $c$, $\| c \|$ is the number of \emph{molecules} of $c$.

For a configuration $c \in \mathbb{N}^\Lambda$ and a reaction $\alpha$ over $\Lambda$, we say that $\alpha = (r,p)$ is \emph{applicable} to $c$ if $r \leq c$. If $\alpha$ is applicable to $c$, then the \emph{result} of applying $\alpha$ to $c$, denoted by $\alpha(c)$, is $c' = c-r+p$. Note that $\alpha(c) \in \mathbb{N}^\Lambda$. In this case, we also write $c \rightarrow_{\alpha} c'$. Moreover, we write $c \rightarrow_{\mathcal{R}} c'$ if $c \rightarrow_{\alpha} c'$ for some reaction $\alpha$ of $\mathcal{R}$. The transitive and reflexive closure of $\rightarrow_{\mathcal{R}}$ is denoted by $\rightarrow_{\mathcal{R}}^*$. We say that $c'$ is \emph{reachable} from $c$ in $\mathcal{R}$ if $c \rightarrow_{\mathcal{R}}^* c'$. If $\mathcal{R}$ is clear from the context, then we simply write $\rightarrow$ and $\rightarrow^*$ for $\rightarrow_{\mathcal{R}}$ and $\rightarrow_{\mathcal{R}}^*$, respectively.

We remark that a CRN is similar to a Petri net $N$ \cite{PetriNet/review/Pet1977} without the initial marking $M$: the set $\Lambda$ corresponds to the set of places of $N$ and the set of reactions $R$ corresponds to the set of transitions of $N$. While in a Petri net distinct transitions in $N$ may correspond to a single reaction in $R$ (i.e., there may be ``copies'' of each transition), this is irrelevant for our purposes.

A CRN is also similar to a vector addition system (VAS) \cite{VASsKarpMiller}. A \emph{VAS} $V$ is a tuple $(\Lambda, S)$ with $\Lambda$ a finite set and $S$ a finite subset of $\mathbb{Z}^\Lambda$. Again, the elements of $\mathbb{N}^\Lambda$ are the \emph{configurations} of $V$. One is interested in the binary relation $\rightarrow$ over $\mathbb{N}^\Lambda$, where $c \rightarrow c'$ if and only if $c' = c + x$ for some $x \in S$. Reachability problems concerning CRNs can be straightforwardly translated to VASs (or Petri nets) and vice versa, see \cite[Appendix A.6]{DBLP:journals/nc/SoloveichikCWB08}.

\subsection{Chemical Reaction Deciders}
A \emph{(leaderless) chemical reaction decider} (CRD, for short) is a tuple $\mathcal{D} = (\Lambda,R,\allowbreak \Sigma,\allowbreak \Upsilon)$, where $(\Lambda,R)$ is a CRN, $\Sigma \subseteq \Lambda$, $\Upsilon: \Lambda \rightarrow \{0,1\}$. The elements of $\Sigma$, $\Upsilon^{-1}(0)$, and $\Upsilon^{-1}(1)$ are called the \emph{input species}, \emph{no voters}, and \emph{yes voters} of $\mathcal{D}$, respectively. Notation and terminology concerning CRNs carry over to CRDs. For example, we may speak of a configuration of $\mathcal{D}$. An \emph{initial configuration} of $\mathcal{D}$ is a nonzero configuration $c$ of $\mathcal{D}$ where $c|_{\Lambda \setminus \Sigma} = 0$ (by abuse of notation we denote the zero vector over suitable alphabet by $0$). A CRD is called \emph{nonincreasing} (\emph{bimolecular}, resp.) if all reactions of $R$ are nonincreasing (bimolecular, resp.).

We define the following function $\Phi_{\mathcal{D}} : \mathbb{N}^\Lambda \rightarrow \{0,1,\und\}$. For $x \in \mathbb{N}^\Lambda$, let $I_x = \{ S \in \Lambda \mid x(S) > 0\}$. Then, for $i \in \{0,1\}$, we have $\Phi_{\mathcal{D}}(x) = i$ if and only if both $I_x \cap \Upsilon^{-1}(i) \neq \emptyset$ and $I_x \cap \Upsilon^{-1}(1-i) = \emptyset$ (as usual, $\Upsilon^{-1}$ denotes the preimage of $\Upsilon$). If $x$ is zero or $I_x \cap \Upsilon^{-1}(0) \neq \emptyset \neq I_x \cap \Upsilon^{-1}(1)$, then $\Phi_{\mathcal{D}}(x) = \und$. Here, the value $\und$ is regarded as ``undefined''.

A configuration $c$ is called \emph{totally stable} (\emph{t-stable} for short) in $\mathcal{D}$ if both $\Phi_{\mathcal{D}}(c) \in \{0,1\}$ and, for all $c'$ with $c \rightarrow^* c'$, we have $c' = c$. Note that if $c$ is t-stable in $\mathcal{D}$, then for all $c'$ with $c \rightarrow c'$, we have $c' = c$. A configuration $c$ is called \emph{output stable} (\emph{o-stable} for short) in $\mathcal{D}$ if both $\Phi_{\mathcal{D}}(c) \in \{0,1\}$ and, for all $c'$ with $c \rightarrow^* c'$, $\Phi_{\mathcal{D}}(c')=\Phi_{\mathcal{D}}(c)$. Note that every t-stable configuration is o-stable. A configuration that is not o-stable (t-stable, resp.) and nonzero is called \emph{o-unstable} (\emph{t-unstable}, resp.).

We say that $\mathcal{D}$ \emph{o-stably decides} (\emph{t-stably decides}, resp.) the function $\varphi: \mathbb{N}^\Sigma \setminus \{0\} \rightarrow \{0,1\}$ if for each initial configuration $c$ of $\mathcal{D}$ and each configuration $c'$ with $c \rightarrow^* c'$, we have $c' \rightarrow^* c''$ where $c''$ is o-stable (t-stable, resp.) in $\mathcal{D}$ and $\varphi(c|_{\Sigma}) = \Phi_{\mathcal{D}}(c'')$. In this case, we also say that $\mathcal{D}$ \emph{o-stably decides} (\emph{t-stably decides}, resp.) the set $\varphi^{-1}(1)$ and that $\mathcal{D}$ is \emph{o-stable} (\emph{t-stable}, resp.).
Note that $\varphi^{-1}(1)$ along with the set $\Sigma$, uniquely determine $\varphi$. In \cite{DBLP:journals/dc/AngluinADFP06} (and \cite{DBLP:conf/dna/DotyH13}), only o-stable CRDs are considered, and as a result the prefix \emph{output} is omitted there.

\begin{remark} \label{remark:no_zero_vec}
We adopt here the definition of o-stably deciding a function/set from \cite[Section~2]{DBLP:conf/podc/AngluinAE06}. In the original definition of o-stably decides from \cite{DBLP:journals/dc/AngluinADFP06}, an initial configuration may be the zero vector and the domain of $\varphi$ contains the zero vector. Since the zero vector corresponds to an input without any molecules and the number of molecules in a bimolecular CRD stays fixed, no molecule can be introduced and, in particular, none of the yes or no voters can be introduced. As a result, there exist no o-stable bimolecular CRDs when (strictly) using the definition of \cite{DBLP:journals/dc/AngluinADFP06}. Finally, we remark that there are (leaderless) CRDs that are o-stable using the definition of \cite{DBLP:journals/dc/AngluinADFP06}, since we may then have reactions $(r,p)$ with $r$ the zero vector. Since $p$ may then be produced at any point in time, an o-stable CRD o-stably decides either $\mathbb{N}^\Sigma$ (when $p$ contains only yes voters) or the empty set (when $p$ contains only no voters). Note that the CRD cannot be o-stable when $p$ contains both yes and no voters. Thus this notion is also not interesting for the (larger) class of CRDs.
\end{remark}

\subsection{Population Protocols}

The notion of population protocol \cite{DBLP:journals/dc/AngluinADFP06,DBLP:journals/eatcs/AspnesR07} is almost equivalent to the notion of bimolecular CRD. The only difference is that, in a population protocol, the set of reactions $R$ is replaced by a \emph{transition function} $\delta: \Lambda^2 \rightarrow \Lambda^2$. In this setting, $\delta(A,B)=(C,D)$ corresponds to the reaction $(r,p)$ with $r = AB$ and $p = CD$ (recall that we may denote vectors by strings). Note that the tuples $(A,B)$ and $(C,D)$ are ordered. Note also that, for given $A,B \in \Lambda$, there are at most two non-mute reactions with $A$ and $B$ as reactants (since we have a transition for $(A,B)$ and for $(B,A)$), while for bimolecular CRDs there can be arbitrary many such reactions.

Reactions, molecules, and species are called \emph{transitions}, \emph{agents}, and \emph{states}, respectively, in the context of population protocols.

An important property of bimolecular CRDs is that the number of molecules stays fixed, i.e., if $c \rightarrow^* c'$, then $\|c\| = \|c'\|$.

\begin{remark}
In \cite{DBLP:journals/dc/AngluinADFP06}, $\delta(A,B)=(C,D)$ is interpreted as follows: a molecule of type $A$ is transformed into a molecule of type $C$ and simultaneously a molecule of type $B$ is transformed into a molecule of type $D$. As a consequence, applying the ``reaction'' $\delta(A,B) = (B,A)$ would result in a different configuration. However, in \cite{DBLP:conf/podc/AngluinAE06} this interpretation is abandoned and $\delta(A,B) = (B,A)$ is considered a mute reaction.
We adopt the convention of \cite{DBLP:conf/podc/AngluinAE06}.
\end{remark}

\section{Semilinearity} \label{sec:semilinear}
In this section we state a number of modest, but useful, observations we made when studying the proof of the semilinearity result of \cite{DBLP:journals/dc/AngluinADFP06}.

Let $\Lambda$ be a finite set. A set $S \subseteq \mathbb{N}^\Lambda$ is called \emph{linear} (over $\Lambda$) if there are $v_0,\ldots,v_n \in \mathbb{N}^\Lambda$ such that $S = \{v_0 + \sum_{i=1}^n k_i v_i \mid k_i \in \mathbb{N}, i \in \{1,\ldots,n\}\}$. A set $S \subseteq \mathbb{N}^\Lambda$ is called \emph{semilinear} (over $\Lambda$) if $S$ is the union of a finite number of linear sets over $\Lambda$.

It is stated in \cite{DBLP:journals/dc/AngluinADFP06} that every semilinear set $S$ is o-stably decidable by a population protocol (i.e., a bimolecular CRD). While this result is often cited in the literature, it is straightforward to verify that the result fails if $S$ contains the zero vector. Indeed, by definition of semilinear sets may contain the zero vector, while the domain of $\varphi$ in the above definition of stably deciding a set is restricted to nonzero vectors (recall from Remark~\ref{remark:no_zero_vec} that we have to use the definition of \cite{DBLP:conf/podc/AngluinAE06} instead of \cite{DBLP:journals/dc/AngluinADFP06}). This small counterexample led us to revisit the proof of \cite{DBLP:journals/dc/AngluinADFP06}. It turns out that Lemma~5 of \cite{DBLP:journals/dc/AngluinADFP06} implicitly assumes that there are at least 2 agents (i.e., molecules), which translate into an initial configuration of size at least 2. Fortunately, this proof can be straightforwardly modified to allow for initial configurations of size 1, by letting, in \cite[Lemma~5]{DBLP:journals/dc/AngluinADFP06}, $I$ map $\sigma_i$ to $(1,b,a_i)$ with $b=1$ if and only if $a_i < c$ for case 1, and with $b=1$ if and only if $a_i = c \mod m$ for case 2 (instead of to $(1,0,a_i)$) --- note that these terms, such as $I$ and $\sigma_i$, are taken from \cite[Lemma~5]{DBLP:journals/dc/AngluinADFP06}. In \cite{DBLP:conf/podc/AngluinAE06} (see also \cite{DBLP:journals/dc/AngluinAER07}), it is shown that if $S \subseteq \mathbb{N}^\Lambda$ is o-stably decidable by a population protocol, then $S$ is semilinear. Thus we have the following (attributed, of course, to \cite{DBLP:journals/dc/AngluinADFP06,DBLP:conf/podc/AngluinAE06}).

\begin{theorem} [\cite{DBLP:journals/dc/AngluinADFP06,DBLP:conf/podc/AngluinAE06}] \label{thm:angl_semilinear}
For every $S \subseteq \mathbb{N}^\Sigma$, $S$ is o-stably decidable by a population protocol (i.e., a bimolecular CRD) if and only if $S$ is both semilinear and does not contain the zero vector.
\end{theorem}

As recalled in \cite{DBLP:conf/dna/ChenDS12}, the result from \cite{DBLP:conf/podc/AngluinAE06} that the sets o-stably decidable by population protocols are semilinear holds not only for population protocols, but for any reflexive and transitive relation $\rightarrow^*$ that respects addition (i.e., for $c,c',x \in \mathbb{N}^\Sigma$, $c \rightarrow^* c'$ implies $c+x \rightarrow^* c' + x$). Hence, Theorem~\ref{thm:angl_semilinear} holds also for the (broader) family of all CRDs.

Another observation one can make when studying \cite{DBLP:journals/dc/AngluinADFP06} is that the proof concerning o-stable CRDs holds unchanged for the smaller class of t-stable CRDs. By expressive power of a family $\mathcal{F}$ of CRDs we mean the family of sets decidable by $\mathcal{F}$. As the result follows from the proof of \cite{DBLP:journals/dc/AngluinADFP06}, we attribute it to \cite{DBLP:journals/dc/AngluinADFP06}.
\begin{theorem} [\cite{DBLP:journals/dc/AngluinADFP06}] \label{thm:char:tot_stable}
The family of t-stable bimolecular CRDs have equal expressive power as the family of o-stable CRDs. Equivalently, the sets that are t-stably decidable by bimolecular CRDs are precisely the semilinear sets without the zero vector.
\end{theorem}
\begin{Proof}
First recall, by the comment below Theorem~\ref{thm:angl_semilinear}, that the expressive powers of the families of o-stable CRDs and o-stable bimolecular CRDs are equal. Now, the family of t-stable bimolecular CRDs is a subset of the family of o-stable bimolecular CRDs. Thus it suffices to show that the if-direction of Theorem~\ref{thm:angl_semilinear} holds for t-stable bimolecular CRDs.

The essential part of the if-direction of the proof of Theorem~\ref{thm:angl_semilinear} above is Lemma 3 and Lemma 5 from \cite{DBLP:journals/dc/AngluinADFP06}. In the proof of Lemma 5 in \cite{DBLP:journals/dc/AngluinADFP06} a population protocol $P$ is described that eventually reaches a configuration $c$ which is called ``stable'' in \cite{DBLP:journals/dc/AngluinADFP06}, and which, in fact, is easily seen to be t-stable (by checking the three conditions of ``stable'' in \cite{DBLP:journals/dc/AngluinADFP06}). The proof of Lemma 3 in \cite{DBLP:journals/dc/AngluinADFP06} trivially holds for t-stable bimolecular CRDs.
\end{Proof}

Since the bimolecular CRDs form a subset of the CRDs, Theorem~\ref{thm:char:tot_stable} holds also when omitting the word ``bimolecular''.

The family of t-stable CRDs form an interesting subclass of CRDs. Indeed, it is easy to verify, during a run of a t-stable CRD, whether or not a configuration is t-stable: one simply needs to verify whether or not there is an applicable (non-mute) reaction. In other words, it is easily verified whether or not the computation has ended. In the larger class of o-stable CRDs, it is not clear whether or not it is computationally easy to verify if a given configuration is o-stable or not. We revisit this latter problem in Section~\ref{sec:det_output_stable}.

The concept of \emph{CRDs with leaders} was introduced in \cite{DBLP:conf/dna/ChenDS12} (it is simply called a CRD in \cite{DBLP:conf/dna/ChenDS12}). The difference with (leaderless) CRDs is that for CRDs with leaders an additional vector $\sigma \in \mathbb{N}^{\Lambda \setminus \Sigma}$ is given and that the initial configurations $c$ have the condition that $c|_{\Lambda \setminus \Sigma}$ is equal to $\sigma$ (instead of equal to $0$). Moreover, in the definition of o/t-stably deciding a function $\varphi$, the domain of $\varphi$ is $\mathbb{N}^\Sigma$ instead of $\mathbb{N}^\Sigma \setminus \{0\}$. Using Theorem~\ref{thm:angl_semilinear}, we now straightforwardly observe that CRDs with leaders decide all semilinear sets. The difference between Theorem~\ref{thm:angl_semilinear} and Theorem~\ref{thm:chen_semilinear} is because the zero vector for (leaderless) CRDs cannot be represented by an initial configuration of positive size, while the zero vector for CRDs with leaders can be represented by an initial configuration of positive size.

\begin{theorem} [\cite{DBLP:conf/dna/ChenDS12}] \label{thm:chen_semilinear}
For every $S \subseteq \mathbb{N}^\Sigma$, $S$ is o-stably decidable by a CRD with leaders if and only if $S$ is semilinear.
\end{theorem}
\begin{Proof}
Again, by \cite{DBLP:conf/podc/AngluinAE06}, every set o-stably decidable by a CRD with leaders is semilinear.

Conversely, let $S \subseteq \mathbb{N}^\Sigma$ be semilinear. Consider $\Sigma' = \{t\} \cup \Sigma$, where $t$ is an element outside $\Sigma$. Let $S' = \{ x \in \mathbb{N}^{\Sigma'} \mid x(t) = 1, x|_{\Sigma} \in S\}$. It is easy to verify that $S'$ is semilinear. Indeed, let $v_0,\ldots,v_n$ be the vectors (cf.\ the definition of linear set) for one of the linear sets that together make up $S$. Then by adding an entry for $t$ with value $1$ for $v_0$ and value $0$ for the other vectors, we see that the obtained vectors define a corresponding linear set for $S'$. Consequently, $S'$ is semilinear. Note that $S'$ does not contain the zero vector. By Theorem~\ref{thm:angl_semilinear}, there is a CRD $\mathcal{D} = (\Lambda, R, \Sigma',\Upsilon)$ that o-stably decides $S'$. Consider now the CRD $\mathcal{D}' = (\Lambda, R, \Sigma,\Upsilon,\sigma)$ with leaders where $\sigma \in \mathbb{N}^{\Lambda \setminus \Sigma}$ is such that $\sigma(t) = 1$ and $\sigma(i) = 0$ if $i \in \Lambda \setminus \Sigma'$. Consequently, the difference between $\mathcal{D}$ and $\mathcal{D}'$ is that index $t$ is not part of the input species. Hence, $\mathcal{D}'$ o-stably decides $S$.
\end{Proof}

Of course, (the proof of) Theorem~\ref{thm:chen_semilinear} also holds by replacing o-stable by t-stable and/or replacing CRDs by bimolecular CRDs.

\section{Determining the output stable configurations} \label{sec:det_output_stable}
In this section we consider the problem of determining whether or not the ``useful'' computation of an o-stable CRD has ended. More precisely, we consider the problem of determining whether or not a given configuration of a o-stable CRD is output stable. Recall from the previous section that it \emph{is} straightforward to determine whether or not a given configuration $c$ is t-stable: one simply needs to check whether or not a non-mute reaction is applicable to $c$ (and check that $\Phi_{\mathcal{D}}(c) \in \{0,1\}$). In Subsection~\ref{ssec:outputS_gen} we consider the o-stable CRDs $\mathcal{D}$ in general and in Subsection~\ref{ssec:outputS_bimol} we consider the case where $\mathcal{D}$ is bimolecular as this subclass turns out to enjoy special properties regarding this problem.

\subsection{The general case} \label{ssec:outputS_gen}

Similar as done in \cite[Theorem~4.2]{DBLP:journals/nc/SoloveichikCWB08}, we formulate now \cite[Corollary~4.1]{VASsKarpMiller} (defined in the context of VASs) in terms of CRNs.
\begin{proposition} [\cite{VASsKarpMiller}] \label{prop:vass_reach_upper}
For given CRN $\mathcal{R}$ and configurations $x$, $y$ of $\mathcal{R}$, it is decidable whether or not $x \rightarrow^* y'$ for some configuration $y' \geq y$.
\end{proposition}

A much more involved result is known as the decidability of the reachability problem for vector addition systems, shown in \cite{VASsMayr} (see \cite{Leroux/VASs/simpleproof} for a simplified proof).
\begin{proposition} [\cite{VASsMayr}] \label{prop:vass_reach_equal}
For given CRN $\mathcal{R}$ and configurations $x$, $y$ of $\mathcal{R}$, it is decidable whether or not $x \rightarrow^* y$.
\end{proposition}
The precise complexity of the reachability problem of Proposition~\ref{prop:vass_reach_equal} is famously unknown (see, e.g., \cite{Leroux/VASs/simpleproof}).

By Propositions~\ref{prop:vass_reach_upper} and \ref{prop:vass_reach_equal} we straightforwardly obtain the following result.
\begin{theorem} \label{prop:outputst_decidable}
For a given o-stable CRD $\mathcal{D}$ and configuration $c$ of $\mathcal{D}$, it is decidable whether or not $c$ is o-stable in $\mathcal{D}$.
\end{theorem}
\begin{Proof}
Testing whether or not $\Phi_{\mathcal{D}}(c) \in \{0,1\}$ is clearly decidable. Let $\Phi_{\mathcal{D}}(c) = j$. Let, for $X \in \Lambda$, $y_X$ be the configuration with $\|y_X\|=1$ and $y_X(X) = 1$. By Proposition~\ref{prop:vass_reach_upper} it is decidable, for each $X \in \Upsilon^{-1}(1-j)$, whether or not there exists a $c'$ such that $c \rightarrow^* c'$ and $c' \geq y_X$, i.e., $c'(X) > 0$. Hence if $c$ contains only yes voters, then we can decide if there is a reachable configuration with no voters (and analogously if $c$ contains only no voters). The only case left to decide is whether or not $c \rightarrow^* 0$ (again, $0$ denotes the zero vector over $\Lambda$). By Proposition~\ref{prop:vass_reach_equal} it is decidable if the zero vector is reachable. Consequently, it is decidable if $c$ is o-stable in $\mathcal{D}$.
\end{Proof}

We now look at properties of o-stable configurations.

Let $\mathcal{D}$ be an o-stable CRD. We now consider the set $U_{\mathcal{D}}$ of all output unstable configurations of $\mathcal{D}$. If $\mathcal{D}$ is clear from the context, then we simply write $U$ for $U_{\mathcal{D}}$. We now recall a useful result from \cite[Lemma~10]{DBLP:conf/podc/AngluinAE06}. For convenience, we also recall its short proof.
\begin{proposition}[\cite{DBLP:conf/podc/AngluinAE06}] \label{prop:unstable_closed}
Let $\mathcal{D}$ be an o-stable CRD. Then $U$ is closed upward under $\leq$. In other words, for all $c,c' \in \mathbb{N}^{\Lambda}$ with $c \leq c'$, if $c \in U$, then $c' \in U$.
\end{proposition}
\begin{Proof}
Let $c \in U$ and $c \leq c'$. If $\Phi_{\mathcal{D}}(c) = \und$, then $c$ contains both yes and no voters (since $c \in U$, $c$ is nonzero). Thus $c'$ also contains both yes and no voters and we have $c' \in U$. Assume that $\Phi_{\mathcal{D}}(c) \in \{0,1\}$. If $\Phi_{\mathcal{D}}(c') = \und$, then there is nothing to prove. Thus assume that $\Phi_{\mathcal{D}}(c) = \Phi_{\mathcal{D}}(c')$. Since $c \in U$, there is a $c''$ with $c \rightarrow^* c''$ with $\Phi_{\mathcal{D}}(c'') \neq \Phi_{\mathcal{D}}(c)$. Let $x := c' - c \in \mathbb{N}^{\Lambda}$. Then $c' = c+x \rightarrow^* c''+x$ with $\Phi_{\mathcal{D}}(c''+x) \neq \Phi_{\mathcal{D}}(c) = \Phi_{\mathcal{D}}(c')$ and $c' \in U$.
\end{Proof}

\begin{remark}
In some papers, such as \cite{DBLP:conf/dna/ChenDS12}, not all species in CRDs need to be voters. In other words, in the definition of CRD we have $\Upsilon: E \rightarrow \{0,1\}$ for some $E \subseteq \Lambda$ (instead of $E = \Lambda$). We remark that Proposition~\ref{prop:unstable_closed} fails in this more general setting. Indeed, if nonzero $c$ contains no voters, then $c \in U$, but by extending $c$ with, say, a yes voter may result in an output stable configuration.
\end{remark}

By Proposition~\ref{prop:unstable_closed}, the set $U$ is characterized by the set $\min(U)$ of minimal elements of $U$ under $\leq$. By Dickson's lemma, recalled below, $\min(U)$ is a finite set.

\begin{proposition}[Dickson's lemma \cite{DicksonsLemma}] \label{prop:dicksons_lemma}
Let $\Lambda$ be a finite set. Then for every $S \subseteq \mathbb{N}^{\Lambda}$, $\min(S)$ is finite.
\end{proposition}

Given an o-stable CRD $\mathcal{D}$ \emph{and} the set $\min(U)$, it is straightforward to verify if a given configuration $c$ is o-stable in $\mathcal{D}$. Indeed, $c$ is o-stable in $\mathcal{D}$ if and only if $u \not \leq c$ for all $u \in \min(U)$. Thus, to check whether or not $c$ is o-stable in $\mathcal{D}$ takes $O(|\min(U)|)$ configuration comparisons, i.e., $O(|\min(U)| \cdot |\Lambda|)$ comparisons of molecule counts. Note that this complexity bound depends only on $\mathcal{D}$, i.e., it is independent of $c$.

This complexity bound can be improved if the vectors of $\min(U)$ are stored in the $k$-fold tree $T_b(k)$ described in \cite{DSkDimQueries/Willard}. In general, given a set $S$ of $n$ vectors of dimension $k$ and vectors $a$ and $b$ of dimension $k$, the topic of orthogonal range querying is concerned with determining the set $X = \{x \in S \mid a \leq x \leq b \}$. A related question is to determine only $|X|$. Using the data structure called $k$-fold trees one can efficiently determine $X$ or $|X|$. In particular, using the $k$-fold tree $T_b(k)$ described in \cite{DSkDimQueries/Willard} it takes $O(\log^{k-\frac{1}{2}}(n))$ vector comparisons to determine whether or not $X = \emptyset$. Now, assume $\min(U)$ is stored in $T_b(k)$. Then a given vector $c$ is o-stable in $\mathcal{D}$ if and only if $X = \{u \in \min(U) \mid u \leq c \}$ is equal to the empty set. Consequently, taking $a$ equal to the zero vector and $b$ equal to $c$, we have the following.

\begin{lemma} \label{lem:ostable_check_kfold}
Given an o-stable CRD $\mathcal{D}$ and assume that the set $\min(U)$ is stored in a $k$-fold tree $T_b(k)$ with $k = |\Lambda|$. Checking whether or not a configuration $c$ is o-stable in $\mathcal{D}$ takes $O(\log^{k-\frac{1}{2}}(n))$ configuration comparisons, with $n = |\min(U)|$.
\end{lemma}
Notice the improvement in complexity of Lemma~\ref{lem:ostable_check_kfold} compared to the straightforward $O(n)$ method mentioned above. Of course, there is a penalty associated with storing $\min(U)$ in a $k$-fold tree: inserting a vector in a $k$-fold tree takes $O(\log^{k-\frac{1}{2}}(n))$ vector comparisons and thus it takes $O(n \log^{k-\frac{1}{2}}(n))$ vector comparisons to set up this data structure. Consequently, the $k$-fold tree method is to be used for testing a (relatively large) \emph{set} of configurations for output stability (instead of just a single configuration $c$).

\subsection{The bimolecular case} \label{ssec:outputS_bimol}

We now show that $\min(U)$ can be efficiently determined when $\mathcal{D}$ is bimolecular. By efficiently determine we mean here that the fraction of time complexity divided by size $|\min(U)|$ of the output is relatively small --- note that $|\min(U)|$ is a naive lower bound on the time complexity. This is particularly useful when one wants to test for o-stability for some large (finite) set of configurations (instead of just a single configuration).

Let, for $k \geq 0$, $\mathcal{C}_{\leq k}$ ($\mathcal{C}_{=k}$, resp.) be the set of configurations $c \in \mathbb{N}^{\Lambda}$ with $\|c\| \leq k$ ($\|c\| =k$, resp.).

We remark that the naive approach to determine whether or not a particular configuration $c$ is o-stable in a o-stable bimolecular CRD $\mathcal{D}$, would compute the set $R_c$ of all configurations reachable from $c$ and then verify that $\Phi_{\mathcal{D}}(c') = \Phi_{\mathcal{D}}(c)$ for all $c' \in R_c$. Note that $R_c \subseteq \mathcal{C}_{=k}$ with $k = \|c\|$ since $\mathcal{D}$ is bimolecular. Thus, in the worst case, one needs to compute in the order of $|\mathcal{C}_{=k}|$ configurations. The value of $|\mathcal{C}_{=k}|$ is equal to the number of multisets of cardinality $k$ over $\Lambda$. This number (called figurate number, simplex number, or multiset coefficient), sometimes denoted by $\multiset{|\Lambda|}{k}$, is equal to the binomial coefficient $\binom{|\Lambda|+k-1}{k}$, see, e.g., \cite[Section~1.2]{EnumComb/RPStanley}.

We start with introducing a new binary relation $\hookrightarrow$. For $c, c' \in \min(U)$, denote $c \hookrightarrow_{\alpha} c'$ if $c \rightarrow_{\alpha} c'+ b$ where $\alpha = (r,p)$ is a reaction and $b$ is some configuration with $b \leq p$ and $b \neq p$. We write $c \hookrightarrow c'$ if $c \hookrightarrow_{\alpha} c'$ for some reaction $\alpha$. It is important to realize that $\hookrightarrow$ is a binary relation on $\min(U)$. Again, the transitive and reflexive closure of $\hookrightarrow$ is denoted by $\hookrightarrow^*$.

We now provide some intuition regarding the notion of $\hookrightarrow$. Intuitively, one may also view $\hookrightarrow$ as a graph $G$ where $\min(U)$ is the set of vertices and $c \hookrightarrow c'$ denotes an arrow from $c$ to $c'$. Lemma~\ref{lem:1step_min_unstable} below proves a number of properties of $\hookrightarrow$. Although Lemma~\ref{lem:1step_min_unstable} does not require that the CRD $\mathcal{D}$ is bimolecular, assume for didactical purposes that $\mathcal{D}$ is bimolecular. Now, Statement~\ref{stmt:go_home} of Lemma~\ref{lem:1step_min_unstable} says that for every vertex of $G$ there is a path in $G$ to a vertex in $M_1 \cup M_2$. The vertices of $M_1 \cup M_2 = M_1 \cup T$ are all of size $2$ and can be readily computed by the first three statements of Lemma~\ref{lem:1step_min_unstable} (because $M_1$ and $T$ can be readily computed). So, to compute all of $\min(U)$ we simply start from $M_1 \cup T$ and obtain all vertices by moving against the arrows. This is accomplished by Algorithm~\ref{alg:gen_min_unstable}. Of course, $G$ is not part of the input and so the arrows of $G$ have to be found dynamically and the candidate vertices have to be checked for membership of $\min(U)$. Statement~\ref{stmt:bimol_decr1} describes the possible differences in size of any two adjacent vertices, which significantly restricts the search space for arrows.

For the next result, recall again that we may denote vectors by strings. Also, note that we do not require in Lemma~\ref{lem:1step_min_unstable} that the CRD $\mathcal{D}$ is bimolecular.
\begin{lemma} \label{lem:1step_min_unstable}
Let $\mathcal{D} = (\Lambda,R,\Sigma,\Upsilon)$ be an o-stable CRD. Let $M_1 = \{ c \in \min(U) \mid \Phi(c) = \und\}$, $M_2 = \{ c \in \min(U) \mid \Phi(c) \in \{0,1\}, c \rightarrow c' \mbox{ for some } c' \mbox{ with } \Phi(c') \neq \Phi(c) \}$, and $T = \{r \in \min(U) \mid (r,p) \in R, \mbox{ and either } \|p\| = 0 \mbox{ or } \Upsilon(A) \neq \Upsilon(B) \mbox{ for some } A,B \in \Lambda \mbox{ with } r(A) \neq 0 \neq p(B) \}$. We have the following.
\begin{enumerate}
\item $M_1 = \{AB \mid A, B \in \Lambda, \Upsilon(A) \neq \Upsilon(B)\}$.

\item $M_2 \subseteq T$.

\item $T \subseteq M_1 \cup M_2$.

\item If $c \rightarrow_{\alpha} c'$ for some $\alpha \in R$, $c \in \min(U)$, and $c' \in U$, then there is a $c'' \in \min(U)$ with $c \hookrightarrow_\alpha c''$.

\item If $\|r\| \geq \|p\|$ for all $(r,p) \in R$, then, for all $c \in \min(U)$, $c \hookrightarrow^* c'$ for some $c' \in M_1 \cup M_2$. \label{stmt:go_home}

\item If $c \hookrightarrow c'$ and $2 = \|r\| \geq \|p\|$ for all $(r,p) \in R$, then $\|c'\| = \|c\|$ or $\|c'\| = \|c\| - 1$. \label{stmt:bimol_decr1}
\end{enumerate}
\end{lemma}
\begin{Proof}
The nonzero configurations where $\Phi(c) = \und$ are those where there are $A, B \in \Lambda$ such that both $c(A) > 0$ and $c(B) > 0$, and $\Upsilon(A) \neq \Upsilon(B)$. The minimal such configurations are such that $c(A) = c(B) = 1$ and $c(X) = 0$ for all other species $X$, and so we obtain the first statement.

We now turn to the second statement. Let $c \in M_2$. Thus $c \in \min(U)$ with $\Phi(c) \in \{0,1\}$ and $c \rightarrow c'$ for some $c'$ with $\Phi(c') \neq \Phi(c)$. Without loss of generality, assume that $\Phi(c)=0$, i.e., $c$ contains only no voters. Let $\alpha = (r,p)$ be the reaction of $\mathcal{D}$ such that $c \rightarrow_{\alpha} c'$. Since $\Phi(c') \neq \Phi(c)$, either $\|p\| = 0$ or a yes voter has been introduced by $\alpha$. As $c \in \min(U)$, we have $c = r$. Also, if a yes voter has been introduced by $\alpha$, then we have $\Upsilon(A) = 0 \neq 1 = \Upsilon(B)$ for some $A,B \in \Lambda$ with $r(A) \neq 0 \neq p(B)$.

We turn to the third statement. Let $\alpha = (r,p)$ be a reaction of $\mathcal{D}$ such that $r \in \min(U)$ and either $\|p\| = 0$ or $\Upsilon(A) \neq \Upsilon(B)$ for some $A,B \in \Lambda$ with $r(A) \neq 0 \neq p(B)$. Assume $r \notin M_1$, i.e., $\Phi(r) \in \{0,1\}$. Then $r \rightarrow_{\alpha} p$ with $\Phi(p) \neq \Phi(r)$ since either $\|p\|=0$ or $\Upsilon(A) \neq \Upsilon(B)$ for some $A,B \in \Lambda$ with $r(A) \neq 0 \neq p(B)$. Consequently, $r \in M_2$.

We now turn to the fourth statement. Let $\alpha = (r,p)$. Since $c \in \min(U)$, we have that $c - r = c' - p \notin U$. Since $c' \in U$, we have $c'' = c' - b \in \min(U)$ for some configuration $b \leq p$ and $b \neq p$. Therefore, $c \hookrightarrow_{\alpha} c''$.

We now turn to the fifth statement. If $c \in M_1$, then we are done. For all $c \in \min(U) \setminus M_1$, $c \rightarrow^* x \rightarrow y$ for some configurations $x$ and $y$ with $\Phi(x) \neq \Phi(y)$. For all such $c$, we assign the value $(k,l)$ where $k = \|c\|$ and $l$ is minimal such that $c \rightarrow^l x \rightarrow y$ for some configurations $x$ and $y$ with $\Phi(x) \neq \Phi(y)$ (by $\rightarrow^l$ we mean the $l$-th power of the relation $\rightarrow$). We show the result by induction on $(k,l)$. If $l = 0$, then $c \in M_2$ and we are done. Assume $l > 0$. Then, by the fourth statement, $c \hookrightarrow c''$ and $c = c'' + b + r - p$. As $\|r\| \geq \|p\|$, we have $\|c''\| \leq \|c\|$. If $\|c''\| < \|c\|$, then, by the induction hypothesis, $c'' \hookrightarrow^* c'$ with $c' \in M_1 \cup M_2$ and so $c \hookrightarrow^* c'$. If $\|c''\| = \|c\|$, then $c'' \rightarrow^{l-1} x \rightarrow y$. This also leads, by the induction hypothesis, to $c'' \hookrightarrow^* c'$ with $c' \in M_1 \cup M_2$ and so $c \hookrightarrow^* c'$.

For the sixth statement, by the definition of the relation $\hookrightarrow$, we have $c'+ b = c - r + p$, $b \leq p$ and $b \neq p$. Thus $c' = c - (r - p + b)$. If $2 = \|r\| \geq \|p\|$, then $1 \leq \|p - b\| \leq \|p\| \leq 2$ and so $0 \leq \|r - p + b\| \leq 1$.
\end{Proof}

Lemma~\ref{lem:1step_min_unstable} above is key for Theorem~\ref{thm:alg_correctness} below. The strategy in the proof of Theorem~\ref{thm:alg_correctness} is to discover all elements of $\min(U)$ ordered by size: first all elements of $\min(U)$ of size $k$ are computed, before any of the elements of $\min(U)$ of size $k+1$ are computed. This ensures that the generated candidates $c$ can be tested for minimality in $U$, i.e., it can be tested whether or not $c \in \min(U)$. Otherwise, the number of generated candidates could potentially grow unbounded.

\begin{theorem} \label{thm:alg_correctness}
Let $\mathcal{D} = (\Lambda,R,\Sigma,\Upsilon)$ be an o-stable bimolecular CRD. Given $\mathcal{D}$, Algorithm~\ref{alg:gen_min_unstable} computes $\min(U)$.
\end{theorem}
\begin{Proof}
First, we initialize $M := M_1 \cup M_2 = M_1 \cup T$ with $M_1$, $M_2$, and $T$ from Lemma~\ref{lem:1step_min_unstable}, see Lines~\ref{ln:init_begin}-\ref{ln:init_end}. Note that the requirement $r \in \min(U)$ in the definition of $T$ is mute as $\mathcal{D}$ is bimolecular: $r \in U$ always holds and since $\mathcal{D}$ is bimolecular, for all $c \in \min(U)$ we have $\|c\| \geq 2 = \|r\|$, and thus $r \in \min(U)$. Also, since $\mathcal{D}$ is bimolecular, the case $\|p\|=0$ in the definition of $T$ is mute. The second (and final) phase is to iteratively augment $M$ with the elements from $\min(U) \setminus (M_1 \cup M_2)$ as prescribed by Statements~\ref{stmt:go_home} and \ref{stmt:bimol_decr1} of Lemma~\ref{lem:1step_min_unstable}.

We show by induction that at Line~\ref{ln:all_minU_until_k}, we have $M_{\mathrm{it}} = \min(U) \cap \mathcal{C}_{=k}$ and $M = \min(U) \cap \mathcal{C}_{\leq k}$.

We first consider the basis case $k=2$. Note that, by Lemma~\ref{lem:1step_min_unstable}, $\min(U) \cap \mathcal{C}_{=2} = \min(U) \cap \mathcal{C}_{\leq 2}$ is obtained from $M_1 \cup M_2$ by adding all $c'$ such that $c' \rightarrow^* c$ and $c \in M_1 \cup M_2$. Note that each such $c'$ is minimal in $U$ as $\|c'\|=2$. This is accomplished in Lines~\ref{ln:it_back_begin}-\ref{ln:it_back_end}.

We now consider the induction step. Let $k \geq 2$. Consider the set $X = \{ c' \mid c' \rightarrow_{\alpha} c + B, \mbox{ for some } \alpha \in R, c \in \min(U) \cap \mathcal{C}_{=k}, B \in \Lambda, c'' \not\leq c' \mbox{ for all } c'' \in \min(U) \cap \mathcal{C}_{\leq k} \}$, where we identify here $B \in \Lambda$ by the configuration $b$ with $\|b\|=1$ and $b(B)=1$. Note that $X \subseteq U$. Since for all $c' \in X$, $\|c'\| = k+1$ and $c' \in U$, we have that $c'' \not\leq c'$ for all $c'' \in \min(U) \cap \mathcal{C}_{\leq k}$ if and only if $c'' \not\leq c'$ for all $c'' \in \min(U)$. Hence $X \subseteq \min(U) \cap \mathcal{C}_{=k+1}$.  The set $X$ is computed in Lines~\ref{ln:begin_new_it}-\ref{ln:innerloop2_end}. Now, by Statements~\ref{stmt:go_home} and \ref{stmt:bimol_decr1} of Lemma~\ref{lem:1step_min_unstable}, $\min(U) \cap \mathcal{C}_{=k+1}$ is obtained from $X$ by adding the configurations $c'$ such that $c' \rightarrow^* c$ with $c \in X$ and $c'' \not\leq c'$ for all $c'' \in \min(U)$. Again, since $\|c'\| = k+1$ and $c' \in U$, we have that $c'' \not\leq c'$ for all $c'' \in \min(U)$ if and only if $c'' \not\leq c'$ for all $c'' \in \min(U) \cap \mathcal{C}_{\leq k}$. These additional configurations $c'$ are (again) computed in Lines~\ref{ln:it_back_begin}-\ref{ln:it_back_end}.

The algorithm halts as by Dickson's Lemma (Proposition~\ref{prop:dicksons_lemma}), $\min(U)$ is finite.
\end{Proof}

\begin{algorithm}[!t]
\begin{algorithmic}[1]
\Procedure{GenMinUnstable}{$\mathcal{D}$}
  \State $T \gets \{r \mid (r,p) \in R, \mbox{ and either } \|p\| = 0 \mbox{ or } \Upsilon(A) \neq \Upsilon(B) \mbox{ for some } A,B \in \Lambda \allowbreak \mbox{ with } \allowbreak r(A) \neq 0 \neq p(B) \}$ \label{ln:init_begin}
  \State $M_{\mathrm{it}} \gets \{AB \mid A, B \in \Lambda, \Upsilon(A) \neq \Upsilon(B)\} \cup T$
  \State $M \gets M_{\mathrm{it}}$
  \label{ln:init_end}
  \While{$M_{\mathrm{it}} \neq \emptyset$}
    \State $M_{\mathrm{new}} \gets M_{\mathrm{it}}$ \label{ln:it_back_begin}
    \While{$M_{\mathrm{new}} \neq \emptyset$}
      \State $M_{\mathrm{old}}, M_{\mathrm{new}} \gets M_{\mathrm{new}}, \emptyset$
      \ForAll{$c \in M_{\mathrm{old}}$, $\alpha \in R$} \label{ln:innerloop1_begin}
        \If{$\exists$ $c'$ with $c' \rightarrow_\alpha c$ and $c'' \not\leq c'$ for all $c'' \in M$}
            \State $M_{\mathrm{new}}, M_{\mathrm{it}}, M \gets M_{\mathrm{new}} \cup \{c'\}, M_{\mathrm{it}} \cup \{c'\}, M \cup \{c'\}$
        \EndIf
      \EndFor \label{ln:innerloop1_end}
    \EndWhile \label{ln:it_back_end}
    \State \label{ln:all_minU_until_k} \Comment{At this point $M = \min(U) \cap \mathcal{C}_{\leq k}$ and $M_{\mathrm{it}} = \min(U) \cap \mathcal{C}_{=k}$.}
    \State $M_{\mathrm{itold}}, M_{\mathrm{it}} \gets M_{\mathrm{it}}, \emptyset$ \label{ln:begin_new_it}
    \ForAll{$c \in M_{\mathrm{itold}}$, $\alpha \in R$, $B \in \Lambda$} \label{ln:innerloop2_begin}
       \If{$\exists$ $c'$ with $c' \rightarrow_\alpha c + B$ and
       $c'' \not\leq c'$ for all $c'' \in M$}
         \State $M_{\mathrm{it}}, M \gets M_{\mathrm{it}} \cup \{c'\}, M \cup \{c'\}$
       \EndIf
    \EndFor \label{ln:innerloop2_end}
  \EndWhile\label{ln:genmin_endwhile}
  \State \textbf{return} $M$
\EndProcedure
\end{algorithmic}
\caption{Generate the set $M$ of minimal output unstable configurations of an o-stable bimolecular CRD $\mathcal{D} = (\Lambda,R,\Sigma,\Upsilon)$}\label{alg:gen_min_unstable}
\end{algorithm}

We now consider the time complexity of Algorithm~\ref{alg:gen_min_unstable}.
\begin{theorem} \label{thm:complexity_alg}
Algorithm~\ref{alg:gen_min_unstable} takes $O(|R| \cdot |\Lambda| \cdot n \log^{|\Lambda|-\frac{1}{2}}(n))$ configuration comparisons to compute $\min(U)$, where $n = |\min(U)|$.
\end{theorem}
\begin{Proof}
There are two inner loops. The first inner loop (at Lines~\ref{ln:innerloop1_begin}-\ref{ln:innerloop1_end}) checks for every $c \in \min(U)$ and $\alpha \in R$, whether or not a $c' \rightarrow_{\alpha} c$ exists, and if such a $c'$ exists, whether or not $c'' \not\leq c'$ for all $c'' \in \min(U) \cap \mathcal{C}_{\leq \|c'\|-1}$. The second inner loop (at Lines~\ref{ln:innerloop2_begin}-\ref{ln:innerloop2_end}) checks for every $c \in \min(U)$, $\alpha \in R$, and $B \in \Lambda$, whether or not a $c' \rightarrow_{\alpha} c + B$ exists, and if such a $c'$ exists, whether or not $c'' \not\leq c'$ for all $c'' \in \min(U) \cap \mathcal{C}_{\leq \|c'\|-1}$. Consequently, the second inner loop is dominant and has at most $n \cdot |R| \cdot |\Lambda|$ iterations. Just like as in Lemma~\ref{lem:ostable_check_kfold} we use the $k$-fold tree $T_b(k)$ from \cite{DSkDimQueries/Willard} to store the vectors of $M$. Again, $k = |\Lambda|$ is the dimension of the vectors of $M$. Recall from Subsection~\ref{ssec:outputS_gen} that it takes $O(\log^{k-\frac{1}{2}}(N))$ configuration comparisons, where $N = |M|$ is the number of elements in $T_b(k)$, to determine if a vector $v$ is such that $w \not\leq v$ for all vectors $w$ in $T_b(k)$. Thus, we require $O(|R| \cdot |\Lambda| \cdot n\log^{|\Lambda|-\frac{1}{2}}(n))$ configuration comparisons. Inserting a vector in $T_b(k)$ takes $O(\log^{k-\frac{1}{2}}(N))$ configuration comparisons and so this step does not dominate. Consequently, we obtain the stated complexity.
\end{Proof}
Note that to obtain the time complexity of Algorithm~\ref{alg:gen_min_unstable} one multiplies the expression of Theorem~\ref{thm:complexity_alg} by the time complexity of comparing two vectors of dimension $|\Lambda|$.

We now obtain the following corollary to Theorem~\ref{thm:complexity_alg} and Lemma~\ref{lem:ostable_check_kfold}.
\begin{corollary}
Given an o-stable bimolecular CRD $\mathcal{D}$, checking o-stability for a set $S$ of configurations using Algorithm~\ref{alg:gen_min_unstable} takes $O((|R| \cdot |\Lambda| \cdot n + |S|) \log^{|\Lambda|-\frac{1}{2}}(n))$ configuration comparisons, with $n = |\min(U)|$.
\end{corollary}

In view of Theorem~\ref{thm:complexity_alg}, it would be interesting to obtain an upper bound on $|\min(U)|$. In fact, it is perhaps reasonable to view $|\min(U)|$ as a measure for the ``complexity'' of the underlying o-stable CRD $\mathcal{D}$. The set $\min(U)$ is an antichain, as any two elements of $\min(U)$ are incomparable (i.e., if $x, y \in \min(U)$ are distinct, then $x \not\leq y$ and $y \not\leq x$). In general, antichains can be arbitrary large for fixed $\Lambda$: for example, for every $k \in \mathbb{N}$, $\mathcal{C}_{=k}$ is an antichain with $|\mathcal{C}_{=k}| = \multiset{|\Lambda|}{k} > k$ if $|\Lambda| \geq 2$. Note however that, by Lemma~\ref{lem:1step_min_unstable}, if $x \in \min(U)$ with $\|x\| = k$, then for every $l \in \{2,\ldots,k-1\}$ there is a $y \in \min(U)$ with $\|y\| = l$. Thus, in particular, $\min(U)$ (for some o-stable bimolecular CRD $\mathcal{D}$) cannot be equal to $\mathcal{C}_{=k}$ for any $k \geq 3$. We expect, but it would be interesting to confirm, that the existence of these ``small'' configurations in $\min(U)$ significantly restricts the cardinality of the antichain $\min(U)$.

In view of Lemma~\ref{lem:1step_min_unstable}, we notice that Algorithm~\ref{alg:gen_min_unstable} works unchanged for the slightly larger class of nonincreasing o-stable CRDs $\mathcal{D}$ with $\|r\| = 2$ for all $(r,p) \in R$. At a significant expense of computational efficiency, Algorithm~\ref{alg:gen_min_unstable} can even be extended to allow for arbitrary nonincreasing o-stable CRDs $\mathcal{D}$. The main issue in extending Algorithm~\ref{alg:gen_min_unstable} to this larger class of CRDs is that Statement~\ref{stmt:bimol_decr1} of Lemma~\ref{lem:1step_min_unstable} is then violated: if $c \hookrightarrow c'$, then $\|c'\|$ can be more than one smaller than $\|c\|$. Indeed, we have $c'+ b = c - r + p$, for some $b \leq p$ with $b \neq p$. Thus $c' = c - (r - p + b)$ and so in the worst case $\|c'\|$ is equal to $\|c\| - (\|r\| - 1)$. As a result, the computational complexity of the dominant second inner loop (at Lines~\ref{ln:innerloop2_begin}-\ref{ln:innerloop2_end}) is significantly increased. Another, smaller issue in extending Algorithm~\ref{alg:gen_min_unstable} is that the elements of the set $T$ may not all be of size $2$. Therefore, the elements of $T$ need to be added to $M$ at the right time (according to their sizes) within the outer while loop instead of at the beginning of the algorithm.

We remark that there is no obvious way to extend Algorithm~\ref{alg:gen_min_unstable} for arbitrary (i.e., also increasing) o-stable CRDs. In particular, it is not clear how to generate the elements of $\min(U)$ in order of their size (as used in the proof of Theorem~\ref{thm:alg_correctness}) since increasing o-stable CRDs may generate large minimal configurations from small minimal configurations. In fact, it is not even clear if it is decidable, given an arbitrary o-stable CRD $\mathcal{D}$ and a finite set $M$ of configurations, whether or not $M = \min(U)$.

\section{Discussion} \label{sec:discussion}
Using the semilinearity proof of \cite{DBLP:journals/dc/AngluinADFP06}, we found that the class of t-stable CRDs have equal expressive power as the larger class of o-stable CRDs. Also, we shown a subtle difference in expressive power between CRDs and CRDs with leaders. Then, we considered the problem of determining whether or not a given configuration $c$ is output stable. In particular, we have shown that the set $\min(U)$ of minimal output unstable configurations may be efficiently computed provided that we restrict to the class of o-stable bimolecular CRDs. Given $\min(U)$ it is straightforward to verify whether or not a given configuration $c$ is output stable.

Various questions regarding the computational complexity of CRDs are open. For example, is it decidable whether or not a given CRD is o-stable, or whether or not it is t-stable? Also, likely some ``bridges'' between the domains of CRDs (functioning as acceptors/deciders) and Petri nets (functioning as generators) remain to be discovered. For example, the semilinear sets are precisely the sets of reachable markings of weakly persistent Petri nets \cite{Yamasaki/PetriNets/Semilinear}. This suggests a possible link between the notions of weak persistence (from the domain of Petri nets) and stable deciders (from the domain of CRDs).

\subsection*{Acknowledgements}
We thank Jan Van den Bussche for interesting discussions on CRNs and for useful comments on an earlier version of this paper. We also thank the anonymous reviewers for useful comments on the paper. R.B.\ is a postdoctoral fellow of the Research Foundation -- Flanders (FWO).

\bibliographystyle{abbrv}
\bibliography{crns_leaderless_outputs_journal}

\end{document}